# The classical mean negative asynchrony in sensorimotor synchronization is not universal in humans. A cross-cultural study.

Julien Lagarde

Julien.lagarde@umontpellier.fr

Euromov DHM

University Montpellier, France

Acknowledgements: The author wish to thank Simone Bastide and Anthony Laurens for help in data collection. Writing of this manuscript was supported by the European Project ENTIMEMENT, H2020-FETPROACT-2018, Grant Number 824160.

Development and learning in interaction with the environment, including repeated exposure and interaction with patterns determined by culture, constitute an example of very slow changes, on an individual's lifespan scale, that influence rhythmic skills (Cameron et al., 20015; Jacoby & McDermott, 2017). Along this line of thinking, we aim at examining how culture pervades across general rhythm skills and specifically determine elementary synchronization.

We used the most elementary expression of human timing function involving movement: The so-called sensorimotor synchronization paradigm. This paradigm has been particularly instrumental in the study of basic human function, with a rare blend of empirical (See Repp, 2005 and Repp & Su, 2013, for reviews; and Repp, 2021 for some historical background), and theoretical- modelling grounded studies (Bose et al., 2019; Chen et al., 1997; Gonzalez, et al., 2019; Hayashi et al., 2016; Kelso et al., 1990; Schöner, 2002). Some of the dynamical and stochastic properties of sensorimotor synchronization, are considered, depending on theoretical assumptions, as belonging to self-organized entrainment phenomena, or based upon explicitly anticipatory processes involving error correction using an interval memory of time intervals. Those properties are often considered fixed in adults, or flexible to some extent and depending on experience (Repp, 2010). For the sake of keeping this manuscript reasonably short we won't review the large literature accumulated on the subject, including behavioural, brain imaging, and modelling.

Our first entry point here is the comparison of Indians and Frenchs participants. Data collected, including 15 French and 15 Indian participants, show interesting differences in the way to synchronize to a simple beat. The data collected points at analysing further in follow ups two time scales of adaptation: Frequency and phase. For definitions and analysis, this study uses the theoretical framework of coordination dynamics. The basic model is a non-linear model of a self-sustained oscillator (l.h.s.), forced by a periodic function and random noise (r.h.s.):

$$\ddot{x} + \dot{x}^3 - \dot{x} + \dot{x}.x^2 + \omega 0 x = \varepsilon.\sin(\omega.t) + \sqrt{Q}.\xi t \qquad Eq.\ 1$$

It is well known that this model of synchronization obeys the so- called theory of Arnold's tongues (Kelso & DeGuzman, 1988), enabling identifying a priori the determiners of synchronization. From this equation relative phase dynamics can be obtained, bistable dynamics of two stable attractors, synchronization and syncopation, resp. in phase and antiphase (Kelso et al., 1990; Eq. 2):

$$\dot{\phi} = \Delta\omega + a\sin\phi - b\sin 2\phi + \sqrt{Q}.\xi t \qquad Eq.2$$

Here we study exclusively synchronization, therefore the bistable equation Eq. 2 can be linearized to obtain further meaningful observables (See Schöner & Kelso, 1986).

We ran an experiment examining the hypothesis that the behavioural difference observed between the Indians and French synchronization comes from sensorimotor adjustments evolving at two time scales, corresponding in short to period or phase adjustments. We aim at i) making this assumption more explicit based on available modelling, and ii) testing explicit predictions from the theory, iii) isolate essential aspects of cultural factors that determine those differences. Among others, we sought to examine rate limits of synchronization and mean phase shift of synchronization in stationary time series. The latter is considered here *equivalent* to the negative mean asynchrony (NMA), known to exhibit "the reported tendency of humans to tap on average prior to tone onsets" (from Bose et al. 2019).

Participants

Indians and French participants (N = 15 in each group, 11 men and 4 women, age 22 to 45), all students at the university, right handed, recruited in Montpellier, were matched in pairs to control for education, age, and musical, or dance, or sports experience. Indians recruited had left India less than 2 years before the experiment, their mother tongue was Indian, their second language English, and they were not fluent in French. Participants gave informed consent before the experiment.

Task

The task was to synchronize as best as possible a tap on the table of the index finger with a sound. 3 trials were completed. The frequency of the sound beats (40ms; carrier frequency 440Hz) sequence was increased every 15 stimuli by 0.3 Hz. The range of the pacing frequency went from 1 to 6.1 Hz.

Data collection

A goniometer was used to collect the index finger position (metacarpophalangeal angle), connected to an A to D card, also used to collect stimuli. To get a good accuracy for determining the temporal center of each auditory beat we collected all signals at 5KHz. A second PC and the sound D to A card was used to display the stimuli.

Data pre-processing

Angular positions were down sampled to 500Hz and low pass filtered at 30Hz with dual pass to negate the phase shift.
Stimuli were processed to identify the time of each center of beat, using a low pass (dual pass) filter and local maxima estimation.

Data processing to measure synchronization

The relative phase between position and beats was estimated taking the value of the Hilbert transform phase of the position at each stimuli onset. Transients (beginning of each plateau) were excluded when calculating mean and dispersion of relative phase. The angular mean and dispersion were estimated and are well defined in the sampled stationary behaviours. The mean relative phase, an angle variable, is here equivalent to the mean time difference, aka mean negative asynchrony (Aschersleben, 2002), classically used in the literature (See Lagarde & Kelso for relations between the two variables).

Results

The maximal rates at which French and Indian participants were able to synchronize were comparable (not shown). However clearly the classical negative mean asynchrony is not observed in the Indians participants (Figure 1).

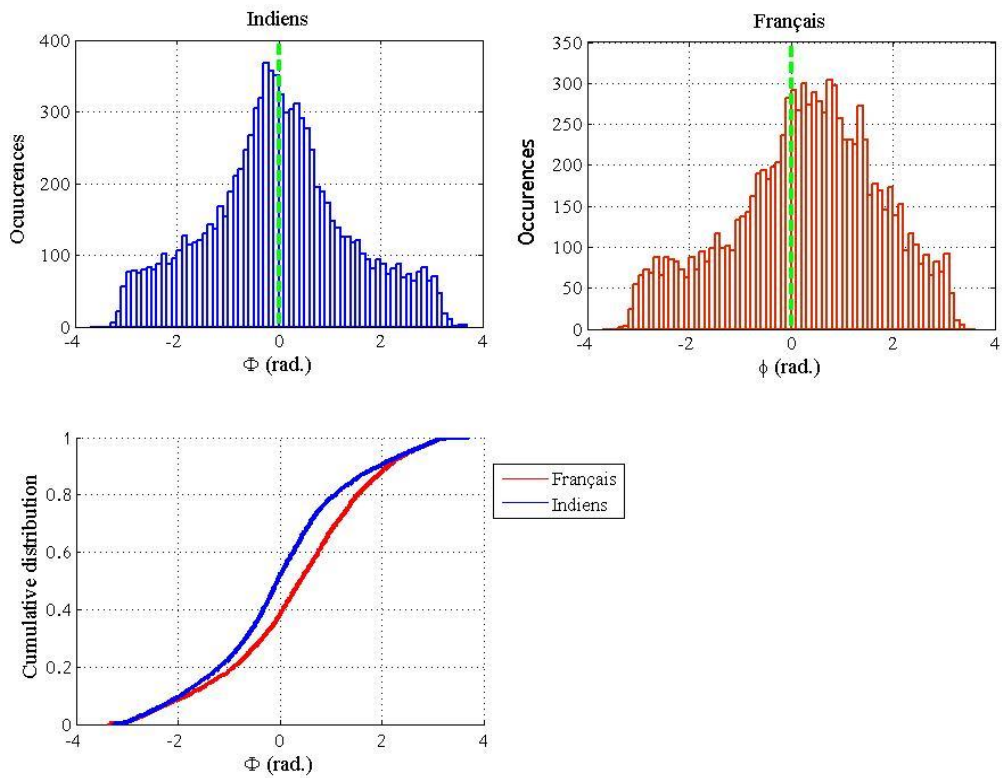

Figure 1. The first task, frequency ranging from 1Hz to 6.1Hz. Histograms of relative phases for all the plateaus for French and Indian participants (N = 9720 values; bin size 0.1 radians). The lower panel shows the cumulative distributions; a Kolmogorov-Smirnov test on the maximal difference between cumulative distributions confirms a significant difference between the distributions of the two groups. The distribution of French participants is centred toward positive relative phase, while for the Indians participants the distribution is centred on negative values. Please remember that the sign is reversed relative to usual conventions: Positive correspond to a movement advance in time with respect to the stimuli, which is the classic mean negative asynchrony.

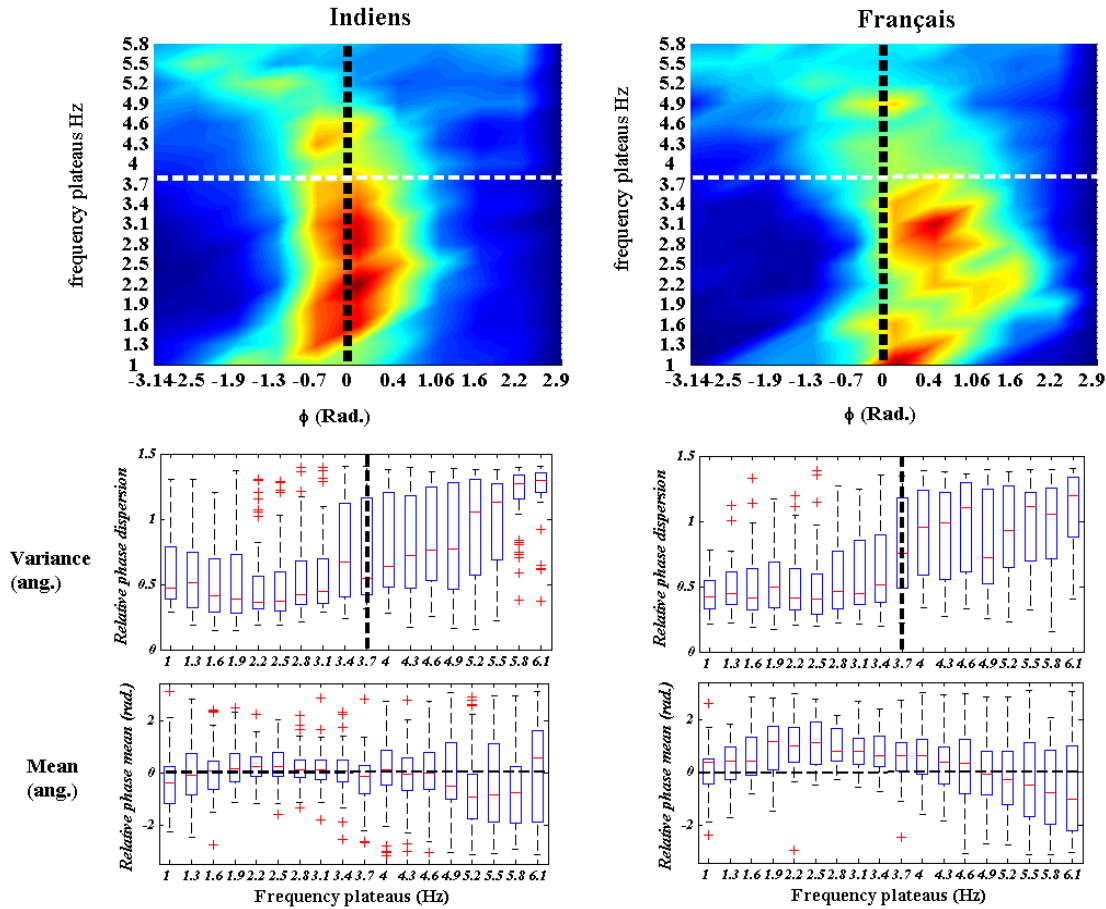

Figure 2 The first task, frequency ranging from 1Hz to 6.1Hz. On the top row the color coded histograms of the relative phase as a function of the frequency of the stimuli (Red is high occurrences, blue is rare occurrences). On the middle row, the box plot of the angular dispersion (variance) of the relative phase. On the bottom row, the box plot of the mean dispersion (variance) of the relative phase. Please remember that the sign is reversed relative to usual conventions: Positive correspond to a movement advance in time with respect to the stimuli, which is the classic mean negative asynchrony.

Discussion

Our hypothesis of an effect of cultural origin on a basic synchronization behaviour was not confirmed by the analysis of the rate limit of synchronization. However, the classical negative mean asynchrony was largely modified by the cultural origin of participants.

The negative mean asynchrony is among the most frequently reported features of the sensorimotor function corresponding to synchronization to a sound in humans (Repp 2005), and has been interpreted as a caused by universal mechanisms, like neurophysiological delays and so called "anticipatory" cognitive processes, driving modeling attempts until recently (See Bose et al., 2019; Ishida & Sawada, 2004). It is considered as a ubiquitous feature of this most basic example of timing function in humans (Aschersleben, 2002). Here we found using the elementary tapping task methodology that this negative mean asynchrony is not universal among humans. This original and unexpected results calls for modelling (See promising perspectives in Ermentrout, 1991), and neuroimaging studies (Jantzen et al., 2004; Nozaradan et al., 2016) from possibly, drastically renewed theoretical assumptions (See Kupferschmidt, 2019).